\documentclass[conference,10pt,letterpaper,nofonttune]{IEEEtran}

\pdfoutput=1
\IEEEoverridecommandlockouts

\usepackage{amsmath,amsfonts}
\usepackage{algorithmic}
\usepackage{algorithm}
\usepackage{array}
\usepackage{dsfont}
\usepackage{enumitem}
\usepackage{textcomp}
\usepackage{stfloats}
\usepackage{url}
\usepackage{verbatim}
\usepackage{graphicx}
\usepackage{xcolor}
\usepackage{soul}
\usepackage[square, numbers, sort&compress]{natbib}
\usepackage{balance}
\usepackage[normalem]{ulem}
\def\BibTeX{{\rm B\kern-.05em{\sc i\kern-.025em b}\kern-.08em
    T\kern-.1667em\lower.7ex\hbox{E}\kern-.125emX}}

\usepackage[scientific-notation=true]{siunitx}

\usepackage{svg}
\usepackage{caption}
\usepackage{subcaption}
\usepackage{mathtools}

\usepackage[font=footnotesize]{subcaption}
\usepackage[font=footnotesize]{caption}

\usepackage[english]{babel}
\usepackage{epsfig}
\usepackage{amscd}
\usepackage[utf8]{inputenc}
\usepackage{mathrsfs}
\usepackage{array}
\graphicspath{ {images/} }
\usepackage{color,listings}
\usepackage{rotating}
\usepackage{acronym}
\usepackage{fancyhdr}
\usepackage{booktabs}
\usepackage[acronyms,nonumberlist,nopostdot,nomain,nogroupskip]{glossaries}

\newacronym{lpwan}{LPWAN}{low-power wide-area network}
\newacronym{sf}{SF}{Spreading Factor}
\newacronym{ml}{ML}{machine learning}
\newacronym{ai}{AI}{Artificial Intelligence}
\newacronym{nbiot}{NB-IoT}{Narrowband IoT}
\newacronym{fl}{FL}{Federated Learning}
\newacronym{arq}{ARQ}{Automatic Repeat Request}
\newacronym{fec}{FEC}{Forward Error Correction}
\newacronym{minlp}{MINLP}{Mixed-Integer Nonlinear Programming}
\newacronym{srn}{SRN}{Standard Radio Node}
\newacronym[plural = Software Defined Radios]{sdr}{SDR}{Software Defined Radio}
\newacronym{mchem}{MCHEM}{Massive Channel Emulator}
\newacronym[plural = Linux Containers]{lxc}{LXC}{Linux Container}

\newacronym{mgen}{MGEN}{Multi-Generator}
\newacronym{kpm}{KPM}{Key Performance Metric}

\newacronym{tgen}{TGEN}{Traffic Generator}
\newacronym[plural = Field Programmable Gate Arrays]{fpga}{FPGA}{Field Programmable Gate Array}
\newacronym{fir}{FIR}{Finite Impulse Response}
\newacronym{rf}{RF}{radio frequency}
\newacronym{dsa}{DSA}{dynamic spectrum access}
\newacronym{rfp}{RFP}{radio fingerprinting}
\newacronym{shd}{SHD}{spectrum hole detection}
\newacronym{fml}{FML}{federated machine learning}
\newacronym{mac}{MAC}{Medium Access Control}
\newacronym{5g}{5G}{fifth generation}
\newacronym{mmw}{mmWave}{millimeter wave}
\newacronym{pus}{PUs}{primary users}
\newacronym{sus}{SUs}{secondary users}

\newacronym{mec}{MEC}{Multi-access Edge Computing}
\newacronym{iot}{IoT}{Internet of Things}
\newacronym{iiot}{IIoT}{Intelligent Internet of Things}
\newacronym{iioit}{IIoT}{Intelligent Internet of Intelligent Things}
\newacronym{drl}{DRL}{Deep Reinforcement Learning}
\newacronym{oran}{O-RAN}{Open Radio Access Network}
\newacronym[plural = Radio Access Networks]{ran}{RAN}{Radio Access Network}
\newacronym[plural = Unmanned Aerial Vehicles]{uav}{UAV}{Unmanned Aerial Vehicle}
\newacronym[plural = Flying ad-hoc Networks]{fanet}{FANET}{Flying ad-hoc Network}
\newacronym[plural = Service Level Agreements]{sla}{SLA}{Service Level Agreement}
\newacronym{zsm}{ZSM}{Zero-touch network and Service Managenent}
\newacronym[plural = Ground Devices]{gd}{GD}{Ground Device}
\newacronym{gdg}{GDG}{Ground Device Group}
\newacronym{qos}{QoS}{Quality of Service}
\newacronym{rl}{RL}{Reinforcement Learning}
\newacronym[plural = Computing Elements]{ce}{CE}{Computing Element}
\newacronym{ric}{RIC}{RAN Intelligent Controller}
\newacronym{etsi}{ETSI}{European Telecommunication Standards Institute}
\newacronym{e2e}{e2e}{end-to-end}
\newacronym{frl}{FRL}{Federated Reinforcement Learning}
\newacronym{tl}{TL}{Transfer Learning}
\newacronym{ddqn}{DDQN}{Double Deep Q-Network}
\newacronym{fo}{FO}{Fanet Orchestrator}

\newacronym{h-home}{H-HOME}{Hierarchical Horizontal Offload ManagEment}
\newacronym{iid}{IID}{Independent and Identically Distributed}

\newacronym{dl}{DL}{Deep Learning}
\newacronym{dnn}{DNN}{deep neural network}
\newacronym{ism}{ISM}{Industrial, Scientific, and Medical}
\newacronym{phy}{PHY}{physical layer}
\newacronym{css}{CSS}{chirp spread spectrum}
\newacronym{crc}{CRC}{cyclic redundancy check}
\newacronym{mno}{MNO}{Mobile Network Operator}
\newacronym{cl}{CL}{Centralized Learning}
\newacronym[plural = Virtual Functions]{vf}{VF}{Virtual Function}
\newacronym{lstm}{LSTM}{Long Short-Term Memory}
\newacronym[plural = Unmanned Vehicles Systems]{uvs}{UVS}{Unmanned Vehicles System}

\newacronym{rsu}{RSU}{Road Side Unit}
\newacronym{obu}{OBU}{On-Board Unit}
\newacronym[plural = M-Boxes]{mbox}{M-Box}{MEC-in-a-box}
\newacronym{mab}{MAB}{Multi-Armed Bandit}
\newacronym{mp-mab}{MP-MAB}{Multi-Player Multi-Armed Bandit}
\newacronym{mp-imab}{MP-IMAB}{Multi-Player Independent MAB}
\newacronym{mp-cmab}{MP-CMAB}{Multi-Player Correlated MAB}
\newacronym{ucb1}{UCB1}{Upper Confidence Bound 1}
\newacronym{urllc}{URLLC}{Ultra-Reliable Low-Latency Communication}
\newacronym{mantra}{MANTRA}{Multi AgeNT Rsu mAnagement}
\newacronym{ztm}{ZTM}{Zero-Touch Management}
\newacronym{ldo}{LDO}{Local Drone Only}
\newacronym{us}{US}{Uniform Selection}
\newacronym{pco}{PCO}{Probabilistic Computation Offloading}

\newacronym[plural = Application Programming Interfaces]{api}{API}{Application Programming Interface}
\newacronym[plural = Telecommunications Operators]{to}{TO}{Telecommunications Operator}

\newacronym{3gpp}{3GPP}{3rd Generation Partnership Project}  

\usepackage{tikz}
\usepackage{tikzscale}
\newif\ifexttikz
\exttikzfalse
\ifexttikz
\else
\usepackage{tikzpagenodes,etoolbox}
\usetikzlibrary{calc}
\usepackage[contents={}]{background}
\AddEverypageHook{%
\ifnumequal{\thepage}{1}{%
    \tikz[remember picture,overlay]{%
        \node[draw,
        minimum width=1.03\textwidth,
        text width=1.02\textwidth,
        font=\footnotesize
        ]
        at ($(current page header area) - (0,5pt)$)
        {%
        This paper has been accepted for publication on International Conference on Computing, Networking and Communications (ICNC 2024). This is the author's accepted version of the article. The final version published by IEEE is R. Raftopoulos, S. D'Oro, T. Melodia and G. Schembra, ``DRL-based Latency-Aware Network Slicing in O-RAN with Time-Varying SLA'' \textit{IEEE ICNC 2024 - IEEE International Conference on Computing, Networking and Communications}, Big Island, Hawaii, USA, 2024.
        };
        \node[draw,
        minimum width=1.03\textwidth,
        text width=1.02\textwidth,
        font=\footnotesize
        ]
        at (current page footer area)
        {%
        ©2024 IEEE. Personal use of this material is permitted. Permission from IEEE must be obtained for all other uses, in any current or future media, including reprinting/republishing this material for advertising or promotional purposes, creating new collective works, for resale or redistribution to servers or lists, or reuse of any copyrighted component of this work in other works.
        };
    }%
}{}
}
\fi

\begin{document}
\markboth{ }%
{Shell \MakeLowercase{\textit{et al.}}: A Sample Article Using IEEEtran.cls for IEEE Journals}


%


\title{DRL-based Latency-Aware Network Slicing in O-RAN with Time-Varying SLAs}


\author{\IEEEauthorblockN{Raoul Raftopoulos\IEEEauthorrefmark{1}, Salvatore D'Oro\IEEEauthorrefmark{2}, Tommaso Melodia\IEEEauthorrefmark{2}, Giovanni Schembra\IEEEauthorrefmark{1}}
\IEEEauthorblockA{
\IEEEauthorrefmark{1}University of Catania, CNIT Research Unit, Catania, Italy\\
\IEEEauthorrefmark{2}Institute for the Wireless Internet of Things, Northeastern University, Boston, MA, U.S.A.\\
E-mail: raoul.raftopoulos@phd.unict.it, giovanni.schembra@unict.it, \{s.doro, melodia\}@northeastern.edu
\thanks{This article is based upon work partially supported by the U.S.\ National Science Foundation under Grant CNS-1925601, by the National Telecommunications and Information Administration (NTIA)'s Public Wireless Supply Chain Innovation Fund (PWSCIF) under Award No. 25-60-IF002, by the European Union under NextGenerationEU PRIN 6GTWINS. 
The work of Raoul Raftopoulos, who has contributed to the design and development of the machine learning algorithm, the setup and execution of the experimental campaign, 
has been supported by the European Union under the Italian National Recovery and Resilience Plan (NRRP) of NextGenerationEU, partnership on “Telecommunications of the Future” (PE0000001 - program “RESTART”, structured project SUPER).
}
}
}

\maketitle

\begin{abstract}
The Open Radio Access Network (Open RAN) paradigm, and its reference architecture proposed by the O-RAN Alliance, is paving the way toward open, interoperable, observable and truly intelligent cellular networks. Crucial to this evolution is Machine Learning (ML), which will play a pivotal role by providing the necessary tools to realize the vision of self-organizing O-RAN systems. However, to be actionable, ML algorithms need to demonstrate high reliability, effectiveness in delivering high performance, and the ability to adapt to varying network conditions, traffic demands and performance requirements. To address these challenges, in this paper we propose a novel Deep Reinforcement Learning (DRL) agent design for O-RAN applications that can learn control policies under varying Service Level Agreement (SLAs) with heterogeneous minimum performance requirements. We focus on the case of RAN slicing and SLAs specifying maximum tolerable end-to-end latency levels. We use the OpenRAN Gym open-source environment to train a DRL agent that can adapt to varying SLAs and compare it against the state-of-the-art. We show that our agent maintains a low SLA violation rate that is $8.3\times$ and $14.4\times$ lower than approaches based on Deep Q-Learning (DQN) and Q-Learning, while consuming respectively $0.3\times$ and $0.6\times$ less resources without the need for re-training.
\end{abstract}


\section{Introduction}

To effectively cater to a wide spectrum of application-specific and user-centric demands, 
there is a need to rethink the way \glspl{ran} are designed and deployed. 
Industry and academia alike commonly agree that this can be achieved by combining three core principles \cite{polese2023understanding}: (i) programmable and virtualized protocol stacks with clearly defined open interfaces; (ii) closed-loop and fast network control and reconfiguration; and (iii) data-driven modeling and \gls{ml}. 
These constitute the foundational principles of the emergent Open RAN paradigm, which has recently garnered substantial momentum as a practical enabler of algorithmic and hardware innovation in future cellular networks \cite{polese2023understanding, zhou2021ran}.
The O-RAN Alliance~\cite{whitepaper2023} is establishing a reference architecture for the Open RAN and defining its specifications with the ultimate goal to push Open RAN toward standardization. In this context, the O-RAN Alliance is promoting open interfaces that connect the various disaggregated functional units, and enable their programmability and monitoring by a set of RAN Intelligent Controllers (RICs). Specifically, O-RAN introduces the Non-real-time (Non-RT) the Near-RT RICs. The former hosts intelligent algorithms called \textit{rApps} that perform inference at timescales larger than $1$s, while the latter hosts \textit{xApps} that operate at timescales from $10$ms to $1$s.  
Intelligent and dynamic network optimization via xApps and rApps introduces novel practical challenges, which include, on one side, designing machine learning (ML) agents capable of adapting to unseen conditions and deployments, and on the other side, selecting features that provide a meaningful representation of the network status without incurring in dimensionality explosion~\cite{nguyen2020deep}.
Moreover, advances in ML-based network automation have been slow, mainly due to the unavailability of large-scale datasets and experimental testing infrastructure to train, test and validate xApps and rApps at scale~\cite{polese2022colo}. 

\textcolor{black}{
In this paper, our goal is to advance the state-of-the-art and provide practical answers to the above challenges and questions by focusing on O-RAN network slicing, a technology necessary to differentiate resource management in the RAN and offer the flexibility, scalability, and adaptability necessary to meet varying and service-specific performance requirements. More specifically, we introduce adaptive and general ML solutions for latency-constrained applications with time-varying Service Level Agreement (SLA).
}


The O-RAN Alliance has identified network slicing as an enabling technology and use case for the O-RAN architecture~\cite{o-rane2smccc}. Several surveys discuss research directions and challenges on network slicing, discussing various aspects, including architecture, orchestration, and resource management \cite{1nets, 2nets}. 
Moreover, the use of xApps to control network slicing in O-RAN has been investigated in several works in the literature~\cite{polese2022colo, rezazadeh2023explanationguided,kouchaki2022actor,nahum2023intent,lotfi2022evolutionary}, which also explore the use of \gls{drl} techniques to regulate slicing and resource allocation policies. 
The authors in \cite{9838462} introduce a framework based on Deep Q-Networks (DQN) for resource allocation and numerology assignment, aiming to meet diverse SLAs. The work presented in \cite{rezazadeh2023explanationguided} addresses a similar challenge but focuses on improving the efficacy of DQN by integrating Explainable Artificial Intelligence (XAI). 
Despite being relevant to our work, both of these methodologies lack explicit formulation and empirical validation on O-RAN deployments. Moreover, they rely upon simulations, thus emphasizing the need for real-world testing and validation. 

Another aspect that remains largely unexplored is the design of DRL agents capable of adapting slicing policies to varying performance requirements. Indeed, while most works in the literature focus on using agents trained to meet certain fixed performance thresholds~\cite{9838462, rezazadeh2023explanationguided}, to the best of our knowledge, there are no works that handle dynamic thresholds and requirements without requiring agent retraining.

To address these key challenges, in this paper, we introduce a novel design of DRL-based xApps for closed-loop control in O-RAN with the objective of computing slicing policies that can satisfy varying SLAs for different network slices while minimizing the amount of Physical Resource Blocks (PRBs) allocated to mobile users, so as to improve resource utilization efficiency and reduce energy consumption. We discuss practical challenges of integrating latency-aware DRL algorithms within O-RAN (e.g., how to generate latency measurements and which interfaces to use to make them available to the RIC), and show that 
our solution can adapt to varying SLAs while improving upon DQN and Q-learning approaches in the literature. Specifically, we show that we can reduce SLA violations by $8.3\times$ and $14.4\times$ while consuming $0.3\times$ to $0.6\times$ less PRBs.

\section{System Model}



We consider a cellular base station (BS) serving a set $\mathcal{U}$ of User Equipments (UEs) and with capacity $C$, i.e., the number of physical resource blocks (PRBs). The BS handles a set $\mathcal{I}$ of network slices and allocates PRBs 
to each network slice dynamically in accordance with their real-time traffic demands and SLAs. 

For each network slice $i \in \mathcal{I}$, the SLA is defined as the 2-tuple $(\Lambda_{i},\varphi^{(\mathrm{SLA})}_i)$.
The SLA specifies a minimum (or maximum) tolerable \gls{kpm} level $\Lambda_{i}$, and a tolerance $\varphi^{(\mathrm{SLA})}_i$ to indicate the percentage of time where $\Lambda_{i}$ must be satisfied. 
Although our design is general, for the sake of simplicity we focus on the case of end-to-end latency where $\Lambda_{i}$ represents the maximum latency level that a certain slice $i$ can tolerate. 
For example, $\varphi^{(\mathrm{SLA})}_i = 0.99$ means that 99\% of the packets must have a latency less than $\Lambda_{i}$. 
We assume that both $\Lambda_{i}$ and $\varphi^{(\mathrm{SLA})}_i$ can vary over time for the same slice.

To satisfy SLAs, network slicing policies are continuously updated so as to meet changing SLAs and determine the number of PRBs that must be assigned to each slice. 
Without loss of generality, we assume that slicing policies are updated on a time-slotted basis. Specifically, we consider $N$ consecutive decision epochs, where each epoch is denoted as $n \in \mathcal{N} = \{1, 2, . . . , N\}$.
The length of each decision epoch depends on the specific application scenario. For example, in the case of xApps, the decision epoch should not exceed $1$s, while it can take larger values in the case of rApps.


\subsection{DRL agent design for adaptive latency-aware slicing}

In this section, we design a DRL agent that continuously monitors network performance and updates slicing policies that satisfy the time-variant SLA for a certain latency-sensitive slice $i \in \mathcal{I}$. 
Specifically, the \textit{action} of the agent corresponds to determining the number of PRBs to dedicate to slice $i$ alone, while the leftover PRBs are allocated to the other slices.
We utilize a variable
$a_{i}(n)$ to indicate this decision with respect to slice $i$ during the $n$-th decision epoch. 

\textcolor{black}{In each decision epoch $n$, we observe the KPM relative to the packets received by the BS. Let $d_{p,i}$ be the value of the end-to-end latency relative to the $p$-th received packet for slice $i$. Let $S_i(n)$ be the total number of received packets during decision epoch $n$. The ratio of the packets whose target latency remains below the maximum tolerable latency level $\Lambda_i(n)$} is:
\begin{equation}\label{eq:k}
    k_i(n) = \frac{1}{S_i(n)} \sum_{p=1}^{S_i(n)} \mathds{1} \left (d_{p, i} < \Lambda_i(n) \right) 
\end{equation}
\noindent 
\textcolor{black}{
where $\mathds{1} (a < b) = 1$ is an indicator function such that $\mathds{1} (a < b) = 1$ if and only if the condition $a<b$ is satisfied, and 0 otherwise. Indeed, $d_{p, i}$ is a function of the action $a_i(n)$. The objective of the agent is to guarantee that the ratio of the packets whose latency exceeds the SLA threshold $k_i(n)$ is less than the SLA tolerance $\varphi^{(\mathrm{SLA})}_i(n)$ while using the minimum number $a_i(n)$ of PRBs. This objective can be formulated as follows: 
}

\begin{align} \label{eq:objective}
    a_i(n) = \arg\min_{a<C} \left\{ \max_a \sum_{n=1}^{N}
    \mathds{1} \left (k_i(n) > \varphi^{(\mathrm{SLA})}_i(n) \right) \right\}
\end{align}
\noindent where $k_i(n)$, defined in \eqref{eq:k}, is a function of the action via the latency function $d_{p, i}$.





The optimization task in \eqref{eq:objective} is challenging as $d_{s,i}$ is hard to model as it depends on the action $a_i(n)$ and other aspects (e.g., traffic load, mobility, modulation) which are hard to capture using closed-form equations. To overcome these challenges, we resort to DRL and its ability to learn the underlying physical models of \glspl{kpm} from the data~\cite{polese2022colo, filali2023communication}. 


 

To compute the observation space of the DRL agent, we consider two important aspects. First, we need to allow the agent to observe relevant KPMs that are correlated to end-to-end latency. Second, we need to maintain the number of such KPMs low because a too high dimensional input may lead to suboptimal performance for ML-driven xApps \cite{Niv8145, inbook}.

Therefore, to solve problem \eqref{eq:objective}, we consider an observation space that includes the number of Transport Blocks (TBs) $tb_{i}$, the ratio of PRB granted and requested $rt_{i}$, and the downlink rate $dl_{i}$, which are highly correlated to the time spent by packets in the transmission buffer~\cite{polese2022colo}. Moreover, since in this paper the objective of the agent is to satisfy latency SLA requirements, we also include the minimum, maximum, and average latency of the packets for all UEs of slice $i$, i.e. $d^{\textrm{min}}_i(n), d^{\textrm{max}}_i(n)$ and $d^{\textrm{mean}}_i(n)$. These metrics are periodically sent from the BSs and UEs to the RIC before a new decision epoch $n$ starts. Section \ref{section: o-ran integration} will discuss more on how these metrics have been collected and processed before being fed to the DRL agent.

\textcolor{black}{Our goal is to develop a DRL agent that learns generalizable policies to meet the SLA requirements even in dynamic scenarios where SLAs vary over time.
For this reason, we add both the current SLA tolerance $\varphi^{(\mathrm{SLA})}_i(n)$ and the maximum tolerable KPM level $\Lambda_i(n)$ to the observation of the agent. This allows our agent to adapt to different SLA requirements without the need to retrain it if the SLA requirements change.
}

Let $\varphi^{\mathrm{(meas)}}_i(n)$ be the measured ratio of packets satisfying the SLA during epoch $n$. The \textit{observation} $o_{i}(n)$ at decision epoch $n$ is defined as:
\begin{equation}
\label{eq: observation}
\begin{split}
o_{i}(n) = \Bigl[ & tb_{i}(n), rt_{i}(n), dl_{i}(n), d^{\textrm{min}}_i(n), d^{\textrm{max}}_i(n), \\
& d^{\textrm{mean}}_i(n), \varphi^{(\mathrm{SLA})}_i(n), \varphi^{\mathrm{(meas)}}_i(n), \Lambda_{i}(n)\Bigr]
\end{split}
\end{equation}


\noindent Finally, 
the \textit{reward} is defined as:

\begin{align}
\label{eq:reward_oran}
r_{i}(n) = & \frac{1}{1 + e^{k \cdot (\varphi^{(\mathrm{SLA})}_i(n) -\varphi^{\mathrm{(meas)}}_i(n))}} \nonumber \\
+ & \left(1 - \frac{a_{i}(n)}{C}\right) \mathds{1}\left(\varphi^{\mathrm{(meas)}}_i(n) \leq \varphi^{(\mathrm{SLA})}_i(n)\right)
\end{align}
\noindent
where $k$ is the sigmoid slope parameter, and we normalize the number $a_{i}(n)$ of PRBs allocated with respect to the capacity $C$.

In (\ref{eq:reward_oran}), we use a sigmoid-like function of the difference between the SLA requirements, $\varphi^{(\mathrm{SLA})}_i(n)$, and the actual ratio of packets that satisfy the SLA, $\varphi^{\mathrm{(meas)}}_i(n)$. This allows the agent to learn how to satisfy the SLA while receiving a higher reward when it is able to use less PRBs. This formulation is helpful as it avoids resource over-provisioning and facilitates energy savings. 


\begin{figure}[t]
    \centering
    \includegraphics[width=0.90\columnwidth]{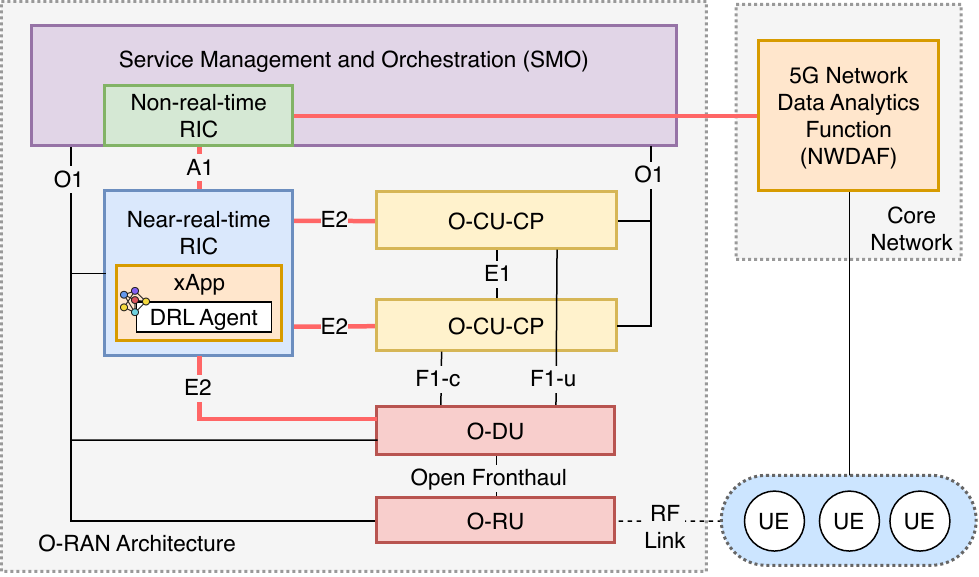}
    \caption{O-RAN architecture with the Integration of the proposed DRL agent.}
    \label{fig: system-architecture oran}
    \vspace{-0.2cm}
\end{figure}

\subsection{Proximal Policy Optimization (PPO) architecture}

We 
implement our solution using the Proximal Policy Optimization (PPO) algorithm \cite{ppo}. PPO is commonly implemented within an Actor-Critic framework, where the policy network functions as the actor, and the value function operates as the critic network. 
PPO features a clipped loss function that adds robustness and stability to the algorithm, making it the state-of-the-art DRL algorithm~\cite{tsampazi2023globecom}.
Specifically, PPO clips the affect of the advantage such that an actor’s action distribution for a particular state does not fluctuate too much during training.



\section{O-RAN Integration and Inference}
\label{section: o-ran integration}

In the following, we provide an overview of the O-RAN architecture and, as shown in Fig. \ref{fig: system-architecture oran}, we illustrate how our DRL agent can be executed as an xApp hosted in the Near-real-time (Near-RT) RIC and how to use O-RAN interfaces (solid lines highlighted in red) to retrieve the data required for its execution. Note that the interface between the 5G network data analytics function (NWDAF) and the Non-RT RIC has not been yet fully specified in O-RAN.


\subsection{O-RAN architecture overview}

As shown in Fig.~\ref{fig: system-architecture oran}, O-RAN embraces RAN disaggregation where base stations (e.g., gNBs) are split into distinct functional units, namely the Radio Unit (RU), Distributed Unit (DU), and Centralized Unit (CU). Each unit is responsible for specific aspects of network functionality, with the RU managing Radio Frequency (RF) components, the DU handling the higher layers of the physical layer, Medium Access Control (MAC), and Radio Link Control (RLC) layers, and the CU overseeing protocol stack layers like Service Data Adaptation Protocol (SDAP), Packet Data Convergence Protocol (PDCP), and Radio Resource Control (RRC). This modular approach allows for independent development, procurement, and operation of the CU, DU, and RU, leading to a more flexible and cost-effective network deployment. Inter-unit interfaces are open and standardized, facilitating RAN telemetry and control exposure to the external environment, ensuring multi-vendor interoperability, and enabling seamless integration of diverse vendors' equipment and solutions into the network.

An important O-RAN innovation is the RIC which offers an abstraction and virtualization layer to support various control loops operating at different timescales, ranging from near-real-time ($10$ms to $1$s) to non-real-time (higher than $1$s). The Near-RT RIC hosts xApps that receive data and telemetry over the E2 interface from the so-called E2 nodes (i.e., DUs, CUs). The E2 interface is logically structured into two distinct protocols: the E2 Application Protocol (E2AP) and E2 Service Model (E2SM). E2AP serves as a procedural protocol, coordinating communication between the near-RT RIC and E2 nodes, while E2SMs, embedded in E2AP messages, implement specific functionalities, such as reporting RAN metrics or controlling RAN parameters. The Non-RT RIC hosts rApps which, similarly to xApps, embed intelligent algorithms to monitor and control the RAN. The Non-RT RIC collects data and control RAN components via the O1 interface, and is connected via the A1 interface to the Near-RT RIC. The A1 interface is used by the Non-RT RIC to exchange enrichment information and policies with the Near-RT RIC. 
The Non-RT RIC is then embedded within the Service Management and Orchestration (SMO) component which handles lifecycle of all O-RAN components.

\begin{figure*}[t]
  \centering
  \begin{subfigure}[t]{0.73\textwidth}
    \includegraphics[width=\linewidth]{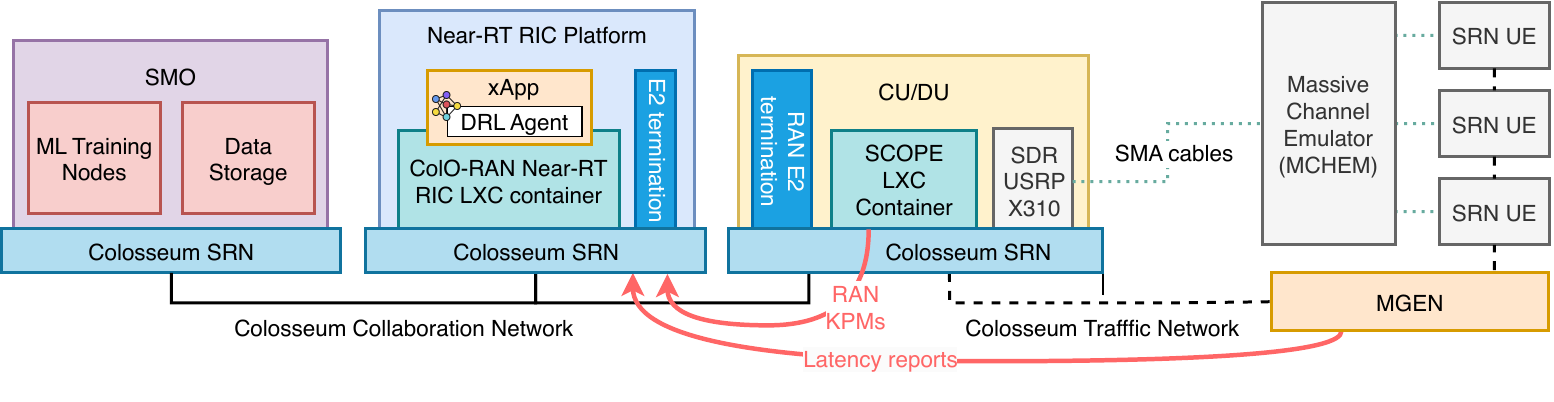}
    \caption{Emulation environment on the Colosseum testbed.}
    \label{fig: non-oran system-architecture oran}
  \end{subfigure}
  \hspace{0.2cm}
  \begin{subfigure}[t]{0.23\textwidth}
  \centering
    \includegraphics[width=\linewidth]{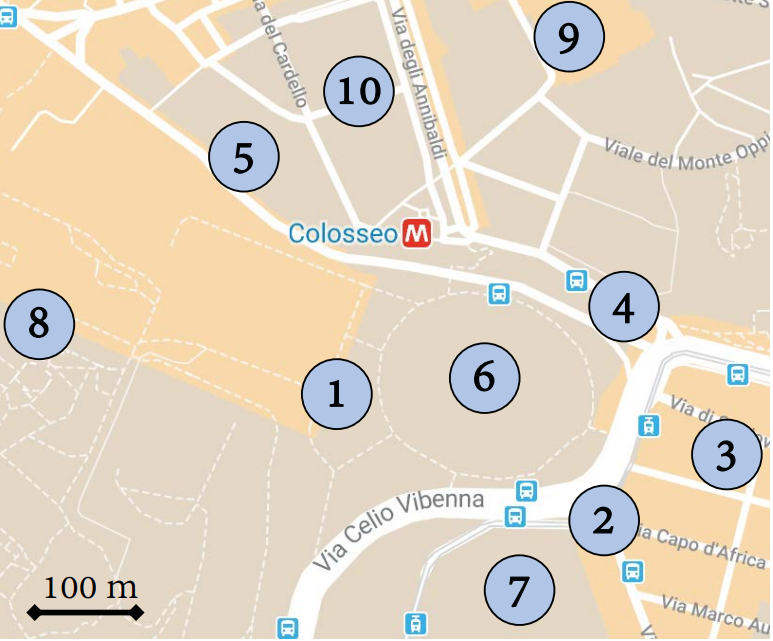}
    \caption{The Rome scenario considered in this paper.}
    \label{fig: rome-bs}
        \end{subfigure}
  
  \caption{O-RAN testbed setup and cellular scenario used for data collection and testing of our DRL agent.}
  \label{fig: results_static_1}
\end{figure*}


\subsection{O-RAN Integration}

Our DRL agent can be executet as an xApp where it can observe the state of the network and take informed decisions to optimize resource management procedures. 
However, it is worth noticing that the Near-RT RIC does not have direct access to end-to-end latency measurements. These are application-layer KPMs that are instead available at the core network. As a consequence, xApps will not be able to obtain such  measurements from E2 nodes directly. For this reason, we leverage the NWDAF which handles data and KPMs at the core network level and makes it available to RAN consumers. As shown in Fig.~\ref{fig: system-architecture oran}, the NWDAF is interfaced with the SMO, which allows the Non-RT RIC to access end-to-end latency measurements for each UE. This data is then shared with the Near-RT RIC via the A1 interface, where it is transmitted as enrichment information, and then made available to the xApps by the Near-RT RIC. The latter is in charge of combining data from the RAN (e.g., CU and DU) collected over the E2 interface with the enrichment information received over the A1 interface. This combination of data is then used to generate the observation in \eqref{eq: observation}, which is then fed to the xApp hosting our DRL agent. The xApp embeds the PPO agent described in the previous section and the following procedures are executed at each decision epoch $n$:
\begin{enumerate}[label=\bfseries Step \arabic*:,leftmargin=*,labelindent=1em]
    \item The Near-RT RIC receives RAN data over E2 (e.g., $tb_{i}(n), rt_{i}(n), dl_{i}(n)$) as well as latency measurements over A1.
    \item Data is fed to the xApp embedding our DRL agent where is it processed and combined to obtain the observation defined in \eqref{eq: observation}. Specifically, individual latency measurements are processed to compute $d^{\textrm{min}}_i(n), d^{\textrm{max}}_i(n)$ and $d^{\textrm{mean}}_i(n)$. Moreover, the xApp also computes the ratio $\varphi^{\mathrm{(meas)}}_i(n)$ of packets that have exceeded the SLA threshold $\Lambda_i$ within the observation period. 
    \item The DRL agent computes a new action (i.e., the number of PRBs to allocate for each slice) and sends it over the E2 interface to the E2 node.
    \item The procedure resumes from Step 1.
\end{enumerate}

\section{Experimental Setup and Data collection}

To generate and collect the data to train and test our DRL agent, we have performed more than 20 hours of experiments (equivalent to more than 1.5 GB of data) on Colosseum, the world's largest Open RAN emulator with hardware in the loop~\cite{colosseum}. We leverage the OpenRAN Gym~\cite{bonati2022openran} open-source O-RAN framework to collect data on Colosseum.
OpenRAN Gym offers access to: (i) a 3GPP-compliant softwarized cellular BS and UE implementations, with support for network slicing, customized scheduling and Physical (PHY)-layer control via the SCOPE framework~\cite{bonati2021scope}, which also embeds data collection pipelines to store more than 30 KPMs in real-time; and (ii) the O-RAN-compliant environment ColO-RAN~\cite{polese2022colo} used to train and test xApps under a variety of deployment scenarios, network and RF conditions across a catalog of predefined cellular scenarios offered by the emulator~\cite{bonati2021scope}.  

The testbed is illustrated in Fig. \ref{fig: non-oran system-architecture oran}. It is worth mentioning that although OpenRAN Gym offers access to RAN-level KPMs via SCOPE, it does not provide an implementation of NWDAF yet. For this reason, we gather core- and application-level KPMs using Colosseum's MGEN (Fig. \ref{fig: non-oran system-architecture oran}) which gives access to end-to-end latency measurements and, thus, offers equivalent functionalities to those offered by NWDAF. Accordingly, in our experiments we leverage SCOPE to instantiate both base stations and UEs and use both MGEN and OpenRAN Gym to collect data to be fed to the DRL agents. Both base stations and UEs are implemented as Linux Containers (LXC) and hosted on Colosseum's Standard Radio Nodes (SRNs)~\cite{colosseum}, which combine a high-performance servers with software-defined radios (SDRs). SDRs generate RF signals that are processed by Colosseum's \gls{mchem}, which is a high-fidelity, large-scale FPGA-based channel emulator. We can use SCOPE to select an RF scenario (a list of cellular scenarios supported in Colosseum is available at \cite{bonati2021scope}). The scenario specifies: (i) how many SRNs can be used in each experiments; and (ii)  a sequence of channel coefficients that describe RF channel conditions between each and every SRNs in the experiment. During the experiment, radios transmit RF signals (cellular waveforms in our case), and \gls{mchem} applies the channel coefficients included in the scenario to them, so as to accurately emulate transmissions happening over the air.

Via SCOPE, OpenRAN Gym supports the concurrent existence of multiple slices on the same BS, and provides APIs to update slicing policies in real-time, so as to compute control strategies that adapt to changing SLA requirements. 
Specifically, by using SCOPE APIs, we can generate PRB masks that are used during the scheduling process to control the number of PRBs for each slice. 
All metrics (both from SCOPE and MGEN) are collected at run-time and saved in CSV format to generate our training dataset. 

In our experiments, we considered the SCOPE Rome urban scenario (Fig. \ref{fig: rome-bs}), in which the locations of the BSs (marked with blue circles) reflect real cell tower deployments extracted from the OpenCelliD database~\cite{opencellid}. The scenario involves a total of 50 nodes: 10 BSs and 40 UEs. Once the experiment begins, Colosseum automatically instantiates both BS and UE containers, emulates RF conditions via its \gls{mchem} and generates traffic via \gls{mgen}.
\gls{mgen} is an open-source traffic generator capable of generating realistic TCP/UDP traffic~\cite{mgen}. It supports a variety of different
classes of traffic with diverse QoS requirements, probability distributions, data rates, and types of service, and generates accurate KPM reports including throughout, jitter and latency.
In our experiments, we consider a uniform traffic profile with a constant bitrate of 1.5 Mbit/s for all UEs. 
We considered a multi-slice network where the agent is required to select how many PRBs to allocate to each slice. The actions taken by the agent are then enforced via SCOPE, which reconfigures the BSs in real-time. The agent observation is generated by periodically reading the dataset entries corresponding to the most recent 250ms of the experiment.

\begin{figure*}[t] 
  \centering
  \begin{minipage}{.45\textwidth}
    \centering
    \includegraphics[width=0.85\columnwidth]{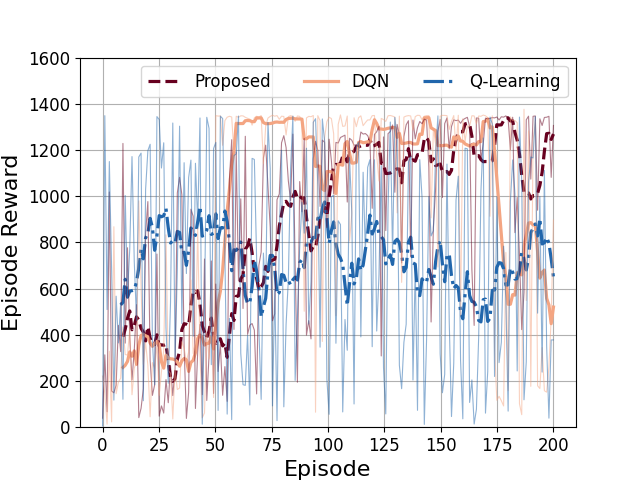}
    \caption{Reward of the first network slice in the first scenario during training ($\Lambda_1$ = 110ms, $\varphi_{1}^{(\mathrm{SLA})} = 0.99$).}
         \label{fig: ep_rew_static_1}
  \end{minipage} \hfill
  \begin{minipage}{.45\textwidth}
    \centering
\includegraphics[width=0.85\columnwidth]{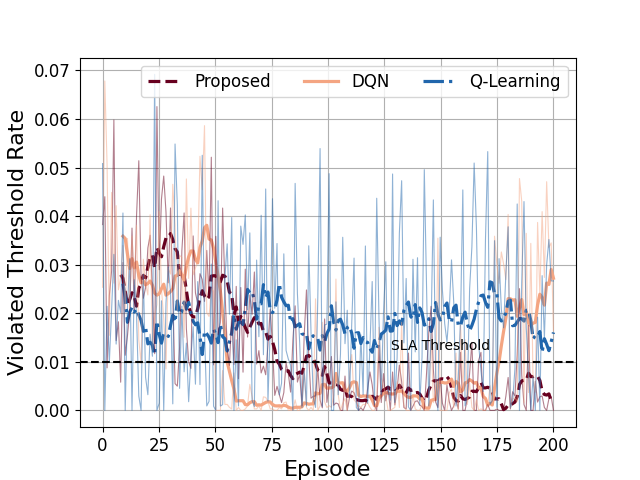}
         \caption{Violation threshold rate of the first network slice in the first scenario during training ($\Lambda_1$ = 110ms, $\varphi_{1}^{(\mathrm{SLA})} = 0.99$).}
         \label{fig: sla_ratio_static_1}
  \end{minipage}\hfill
    \begin{minipage}{.45\textwidth}
    \centering
    \includegraphics[width=0.85\columnwidth]{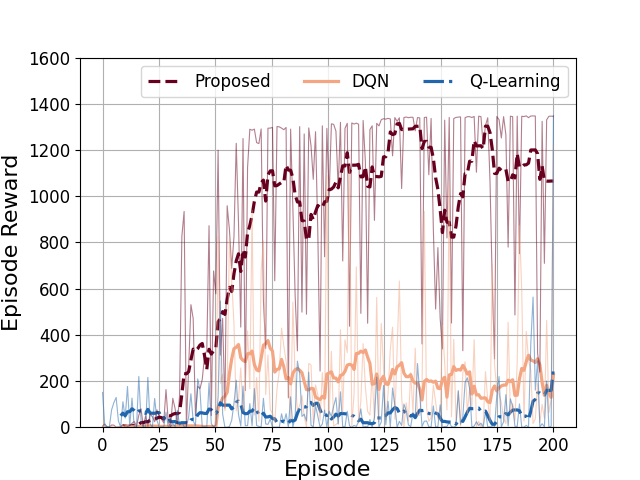}
         \caption{Reward of the second network slice in the first scenario during training ($\Lambda_2 = 50$ms, $\varphi_{2}^{(\mathrm{SLA})} = 0.99$).}
         \label{fig: ep_rew_static_2}
  \end{minipage}\hfill
  \begin{minipage}{.45\textwidth}
    \centering
    \includegraphics[width=0.85\columnwidth]{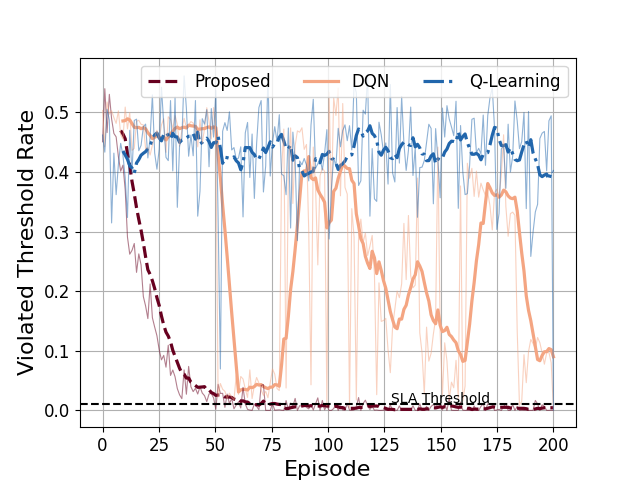}
         \caption{Violation threshold rate of the second network slice in the first scenario during training ($\Lambda_2 = 50$ms, $\varphi_{2}^{(\mathrm{SLA})} = 0.99$).}
         \label{fig: sla_ratio_static_2}
  \end{minipage}
\end{figure*}

\textcolor{black}{Following O-RAN specifications~\cite{polese2023understanding}, we perform training offline,
which involves training a model using datasets collected in advance, rather than interacting with the environment in real-time. The dataset consists of tuples $(s, a, r, s')$ where $s$ is the current state, $a$ is the action taken in that state, $r$ is the immediate reward received, and $s'$ is the next state after taking action $a$ in state $s$. The training starts with an initial state-action pair, where $s_0$ is the initial state, and $a_0$ is an action to explore. Since the training happens offline on static data, it is hard to maintain temporal relationships between state transitions (which would require a combinatorial number of state transition pairs). For this reason, we use a more practical approach where the next state $s_1$ is randomly sampled from the dataset among all of the instances in which action $a_0$ has been taken in state $s_0$. Although this approach does not maintain temporal relationships between state transitions, on average, it guarantees that the statistical behavior of state transitions is captured given that the number of explored states is large enough.}

\section{Numerical Results}
\begin{figure*}[t] 
  \centering
  \begin{minipage}{.45\textwidth}
    \centering
    \includegraphics[width=0.85\columnwidth]{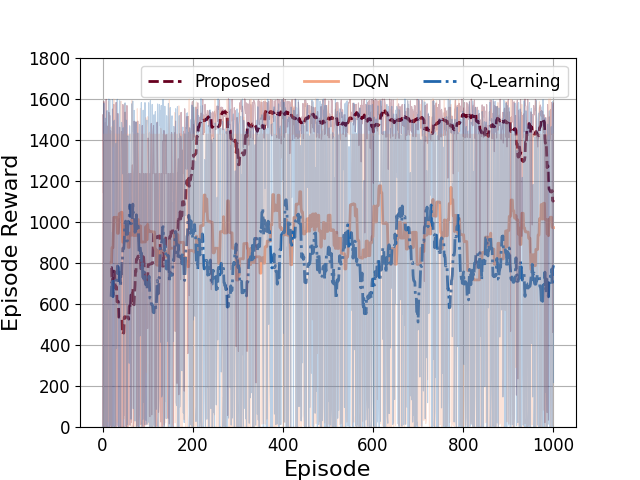}
         \caption{Reward in Scenario DYN during training.}
         \label{fig: ep_rew_dynamic}
  \end{minipage}\hfill
  \begin{minipage}{.45\textwidth}
    \centering
     \includegraphics[width=0.85\columnwidth]{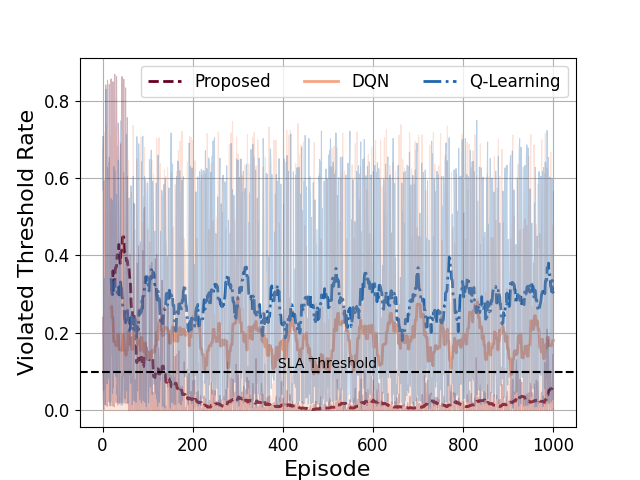}
         \caption{Violation threshold rate in Scenario DYN during training.}
         \label{fig: cdf}
  \end{minipage}\hfill
 
\end{figure*}


We have evaluated our DRL-based approach across the following two scenarios:

\begin{itemize}
\item \textbf{STAT Scenario:}~In this first scenario, we consider two network slices where SLA requirements for each slice do not change over time. We set $\Lambda_1 = 110ms$ and $\Lambda_2 = 50ms$ for the first and second slice, respectively. We also fix $\varphi_{1}^{(\mathrm{SLA})} = \varphi_{2}^{(\mathrm{SLA})} = 0.99$. The objective of this analysis is to show that our approach improves upon the literature even in the case of static SLAs~\cite{9838462, rezazadeh2023explanationguided}. 
\item \textbf{DYN Scenario:}~In this second scenario, we consider one network slice and a dynamic scenario where the SLA is generated at random at each episode. Specifically, at the start of each episode, we select at random the maximum tolerated latency value $\Lambda_1$ such that $\Lambda_1 \in [10, 110]ms$. Moreover, we periodically evaluate the agents' performance for two specific SLA profiles, i.e., for $\Lambda_1 =110ms$ and $\Lambda_1 =30ms$. $\varphi_{1}^{(\mathrm{SLA})}$ is constant and equal to $\varphi_{1}^{(\mathrm{SLA})} = 0.9$. This second scenario highlights the adaptability of our approach and its effectiveness in satisfying diverse SLA levels without the need to retrain the DRL agent.
\end{itemize}

We compare to similar DQN-based approaches such as those in \cite{9838462, rezazadeh2023explanationguided}. Specifically, the work presented in \cite{9838462} introduces a DQN-based framework for resource allocation and numerology assignment to satisfy diverse SLAs, while the authors in \cite{rezazadeh2023explanationguided} focus on a similar problem but integrate Explainable Artificial Intelligence (XAI) to improve DQN efficacy. 
Furthermore, we also compare against conventional Q-Learning in an effort to evaluate the benefits of using Deep Neural Networks (DNNs) for complex control problems with dynamic parameters.

The training time for each agent has requested approximately 13 hours on an NVIDIA V100 with 32GB of memory. 

\subsection{STAT Scenario: Performance evaluation and comparison}

In the STAT Scenario, we consider static SLAs for two slices with $\Lambda_1 = 110ms$ and $\Lambda_2 = 50ms$, and $\varphi_{1}^{(\mathrm{SLA})} = \varphi_{2}^{(\mathrm{SLA})} = 0.99$.
Figs. \ref{fig: ep_rew_static_1} and \ref{fig: sla_ratio_static_1} illustrates the training phase under STAT Scenario for the first network slice. 
Specifically, we showcase the episode reward (Fig. \ref{fig: ep_rew_static_1}) and the ratio of packets violating the latency threshold $\varphi_{sla}$ stipulated by the SLAs (Fig. \ref{fig: sla_ratio_static_1}). 
Black dashed lines are used to identify the maximum tolerable SLA violation rate (i.e., $1-\varphi_{1}^{(\mathrm{SLA})}$ and $1-\varphi_{2}^{(\mathrm{SLA})}$). 
From Fig. \ref{fig: ep_rew_static_1}, we notice that our approach converges to a stable policy that satisfies both $\Lambda_1 = 110ms$ and $\varphi_{1}^{(\mathrm{SLA})}=0.99$ within 150 episodes. 
The Q-Learning approach fails to converge and consistently violates SLAs.
On the contrary, the DQN-based solutions converge to a good policy at first, which is then forgotten at around episode 175 and replaced by a less efficient policy that violates SLAs.



Regarding the second network slice, and as shown in Figs. \ref{fig: ep_rew_static_2} and \ref{fig: sla_ratio_static_2}, we notice that the Q-Learning approach still fails to yield a satisfactory solution. DRL-based solutions exhibit better performance if compared to Q-Learning, but consistently fail in satisfying SLAs. Our proposed approach stands out as the sole method capable of fulfilling these requirements. This distinction is further corroborated by the episode reward, which averages 4.5$\times$ and 12$\times$ higher than the DQN-based and Q-learning-based solutions, respectively.

\subsection{DYN Scenario: performance evaluation and comparison}

In the DYN Scenario, we evaluate the adaptability of our DRL design against varying SLA levels. We consider a single slice where the maximum tolerable latency $\Lambda_1$ is modeled as a random variable whose value is selected in the range $[10, 110]ms$ at random at the beginning of each training episode. This phase is instrumental in enabling our model to learn how to adapt and satisfy different SLAs under diverse conditions. Moreover, we conduct periodic evaluation of the effectiveness of learned policies. Due to space limitations, we only consider two evaluation configurations with $\Lambda_1\in\{30,110\}$.

\begin{figure*}[t]
 \begin{minipage}{.33\textwidth}
    \centering
      \includegraphics[width=\linewidth]{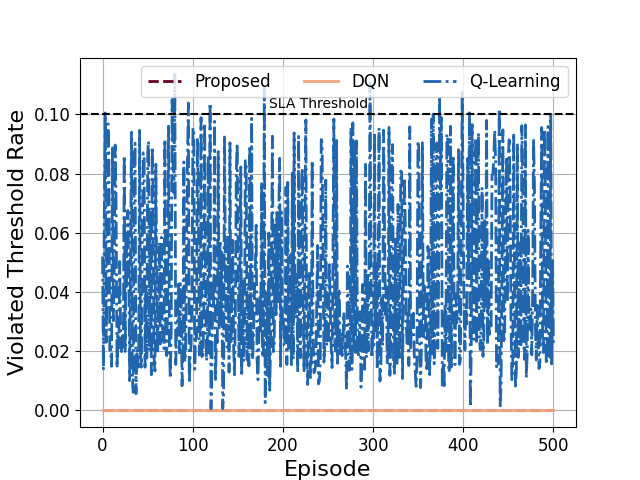}
         \caption{Violation threshold rate for $\Lambda_1 = 110$ms and $\varphi_{1}^{(\mathrm{SLA})} = 0.90$.}
         \label{fig: sla_ratio_dynamic_1}
  \end{minipage}\hfill
  \begin{minipage}{.33\textwidth}
    \centering
     \includegraphics[width=\linewidth]{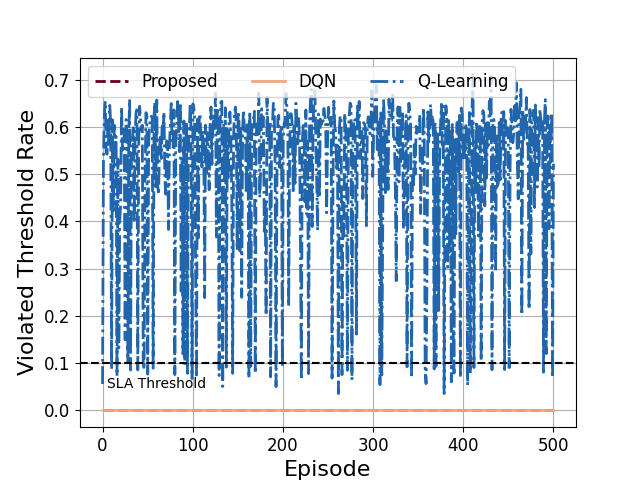}
         \caption{Violated threshold rate for $\Lambda_1 = 30$ms and $\varphi_{1}^{(\mathrm{SLA})} = 0.90$.}
         \label{fig: sla_ratio_dynamic_2}
  \end{minipage}\hfill
  \begin{minipage}{.33\textwidth}
    \centering
     \includegraphics[width=\linewidth]{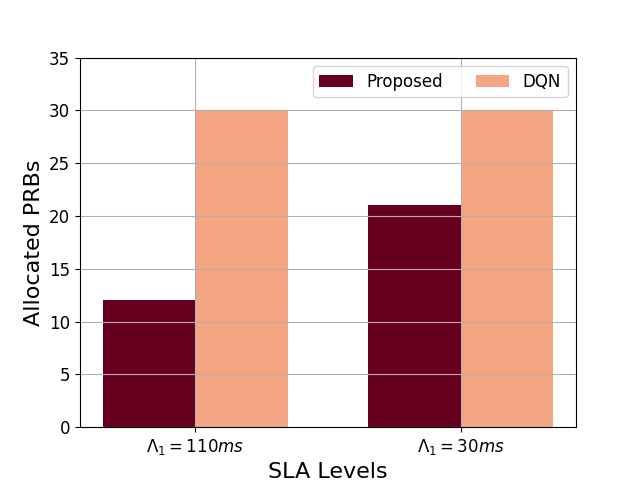}
         \caption{Average number of allocated PRBs for $\Lambda_1 = 30$ms and $\varphi_{1}^{(\mathrm{SLA})} = 0.90$.}
         \label{fig: usedprbs}
  \end{minipage}
\end{figure*}

In Figs. \ref{fig: ep_rew_dynamic} and \ref{fig: cdf}, we show the reward and violation rate of different agents during training in the DYN Scenario. Fig. \ref{fig: ep_rew_dynamic} shows that our proposed approach outperforms the DQN-based and Q-Learning approaches by delivering a higher reward. Similarly, Fig.~\ref{fig: cdf} shows that our approach also delivers the lowers violation rate.
This is, on average, $8.3\times$ and $14.4\times$ lower than DQN and Q-Learning, respectively.


During training, we also evaluate the effectiveness of the learned policies over time. Figs. \ref{fig: sla_ratio_dynamic_1} and  \ref{fig: sla_ratio_dynamic_2} show that both our proposed methodology and the DQN-based approaches effectively meet the SLA requirements. However, the Q-Learning satisfies SLAs only when $\Lambda_1 = 110$ms (Fig.~\ref{fig: sla_ratio_dynamic_1}). 

We further investigate resource utilization aspects for the different agents. Since Q-Learning fails in satisfying SLAs, we only report data for our proposed and DQN-based approaches. Fig. \ref{fig: usedprbs} shows how many PRBs on average are allocated when $\Lambda_1 \in \{30,110\}$ms. We see that the DQN-based approach allocates always 30 PRBs, while our approach allocates respectively 60\% and 30\% less PRBs in the case of $\Lambda_1 = 110$ms and $\Lambda_1 = 30$ms, respectively. These results show that our approach is able to adapt better to SLA variations and learns to allocate fewer PRBs to save energy. 
\section{Conclusions and Future Work}

In this paper, we have presented the design of a DRL agent that can learn to react to varying SLAs without requiring any retraining. We design the DRL agent in a way that makes it suitable for O-RAN applications and illustrate the integration efforts required to execute our agent as an xApp hosted in the Near-RT RIC. We train our agent using a PPO architecture on data collected using OpenRAN Gym and Colosseum, and show that our solution can adapt to varying SLAs while improving upon DQN and Q-learning approaches in the literature. Specifically, we show that we can reduce \textcolor{black}{SLA violations by $8.3\times$ and $14.4\times$ while consuming from $0.3\times$ to $0.6\times$ less PRBs.} In the future, we will investigate the case of multi-agent DRL solutions coordinating decisions to simultaneously satisfy diverse SLAs, with possibly different target KPMs.



\balance
\footnotesize
\bibliographystyle{IEEEtran}
\bibliography{bibliography.bib}

\vfill

\end{document}